\documentclass[showpacs,amsmath,amssymb,preprint]{revtex4}
\usepackage{amsmath}    % need for subequations
\usepackage{graphicx}   % need for figures
\usepackage{epsfig}   % need for figures
\usepackage{verbatim}   % useful for program listings
\usepackage{color}      % use if color is used in text
\usepackage{subfigure}  % use for side-by-side figures
\pdfoutput=1

\begin{document}

\title{Amorphous to amorphous transition in particle rafts }
\author{Atul Varshney, A. Sane, Shankar Ghosh and S. Bhattacharya}
\affiliation{Department of Condensed Matter Physics and Materials Science,~ Tata Institute of Fundamental Research,~ Homi Bhabha Road,~ Mumbai 400-005,~ India\\
}
%\pacs{46.55.+d}{}
%\pacs{nn.mm.xx}{Second pacs description}
%\pacs{nn.mm.xx}{Third pacs description}

\begin{abstract}

Space-filling assemblies of athermal hydrophobic particles floating at an air-water interface, called particle rafts, are shown to undergo an unusual phase transition  between two  i.e., a low density `less-rigid and a high density `more-rigid'  amorphous states as a function of particulate number density ($\Phi$).
The former is shown to be a capillary-bridged solid and the later a  frictionally-coupled  one. Simultaneous studies involving direct imaging  as well as measuring its  mechanical response to longitudinal and shear stresses show that   the transition  is marked by a subtle structural anomaly  and a  weakening of the shear response. The structural anomaly is identified from the variation of the
mean  coordination number, mean area of the Voronoi cells and the  spatial profile of the displacement field with $\Phi$. The weakened shear response is related to local plastic instabilities caused by the depinning of the contact-line of the underlying fluid on the rough surfaces of the particles.

\end{abstract}
\pacs{46.55.+d}
\maketitle

\section{Introduction}

In crystalline solids in thermodynamic equilibrium, the onset of rigidity is a direct consequence of the appearance of long range positional order and the broken continuous translational symmetry. No such overarching principles are known to govern amorphous solids which are frequently out of equilibrium. For example, the onset of rigidity in athermal granular systems  has been a topic of great recent interest \cite{liu_jamming_2010}.
Such studies explore the transition of a  system from a state of zero rigidity  (unjammed state) to that of a finite one (jammed state) \cite{liu_nonlinear_1998,bi_jamming_2011,durian_foam_1995,zhang_jamming_2005}.  But only a few have explored  further phase transitions that may exist within the jammed state of a system \cite{zhao_new_2011}.   In this paper we  study the mechanical response of   a rigid but amorphous particle raft  to   compressive (longitudinal) and shear (transverse) stresses for varying  particulate number density ($\Phi$).  These  rafts are  space-filling assemblies of athermal hydrophobic particles floating at an air-water interface and have properties  common to elastic  \cite{vella_elasticity_2004} and granular \cite{cicuta_granular_2009} solids.  Its   mechanical  response  and structural reorganization reveal  anomalies  that are suggestive of a  phase transition between two amorphous  states, i.e.,   a low-density `less-rigid'  state  and  a high-density `more-rigid' state.

\section{Experimental Details}
\subsection{Particle raft:  A rigid and athermal model system}
Surface tension assists hydrophobic particles which are denser than water to float on it. They deform the otherwise flat liquid surface
\cite{rapacchietta_force_1977,fournier_anisotropic_2002}  under gravity,~ thereby generating long-range inter-particle attraction and form particulate-clusters \cite{berhanu_heterogeneous_2010,chan_interaction_1981}.  The individual clusters show solid like properties,~ i.e.,~ they retain their shape and have a finite rigidity. However,~ for small areal coverage,  these clusters are sparsely  distributed.  Hence,  at length scales comparable to the system size the collective mechanical  property of the floating    clusters is governed by the intervening liquid. Upon compression, i.e., increasing the areal coverage,~  the individual clusters  fuse to form a contiguous system  spanning structure. The short-range  inter-particle interactions  depend  on the surface roughness and wettability of the particles  and are either  attractive or repulsive \cite{chan_interaction_1981,kralchevsky_particles_2001}.  The poly-dispersity and the athermal nature of the particles make the raft  amorphous.

\subsection{ Experimental protocol to prepare the particle raft  and its structural characterisation}

\begin{figure}
\includegraphics[width=0.75\textwidth]{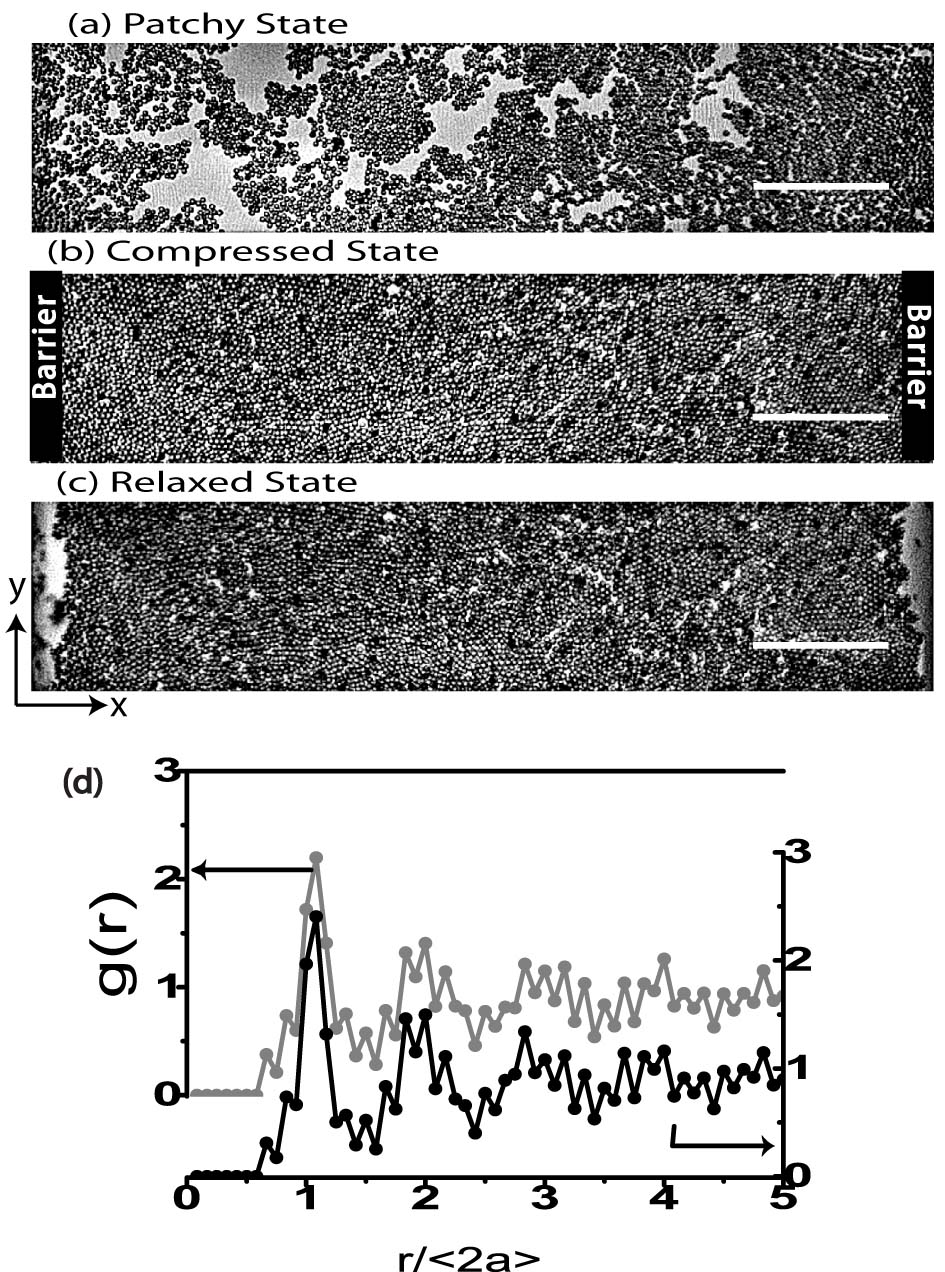}

\caption{Micrographs of hydrophobized silica particles at the air-water interface in the form of (a) disjoint clusters (patchy state) and (b) a homogeneous `compressed' state of the raft in presence and (c) absence ('relaxed' state) of  longitudinal stress. The scale bars, 10mm. (d) The left (right) panel shows the plot of the radial density pair correlation function,~ $g(r)$,~ of the `relaxed' (`compressed') state of the system.  }
\label{fig.1}
\end{figure}

The following  experimental protocol is used to prepare the  space filling structure of the  raft.\\
(i) The hydrophobic (coated with FluoroPel PFC M1104V-FS from Cytonix LLC) silica particles of average diameter ($2a$)  are initially sprinkled on the air-water interface in a Langmuir trough. These particles form disjoint  particulate clusters (see Fig.\ref{fig.1}(a)). This  state of the system is defined as  a ` patchy' state. For  data presented in this paper,~ unless otherwise mentioned,~ $2a$ =0.5mm (polydispersity is  15$\%$ and density of silica  is 2500 $Kg/m^3$).\\
(ii) These  clusters are then  brought within the interaction distance, i.e.,  capillary length ($L_c \sim 2 mm $), \cite{berhanu_heterogeneous_2010,capillary}~ by moving the motorized Teflon barriers of the trough inwards. This constitutes the  first compression cycle ($c1$). As a result of this compression the
disjoint clusters coalesce to form a system-spanning
quasi-two dimensional raft  \cite{kralchevsky_particles_2001,vella_elasticity_2004,cicuta_granular_2009}.   The inward motion of the Teflon barriers is  stopped just before the out-of-plane deformations (wrinkling) \cite{vella_elasticity_2004} of the raft appear. The resulting   homogeneous `compressed'  state of the particle raft is shown in Fig. \ref{fig.1}(b).\\
(iii) The barriers are  then moved outward in the  first expansion cycle ($e1$)  until they detach from the raft (see movie `MS1.avi' in Supplementary information). Figure \ref{fig.1}(c) shows the micrograph of the resulting relaxed,~ yet rigid, state of the raft.   Further cycles of compression and expansion transform the system between the `compressed' and the `relaxed' states.

The  number density of the particles ($\Phi$) between the barriers  defined as,~ $\Phi=N\pi a^2/(L_x L_y)$,~ where $L_x$ and $L_y$ (=140mm) are the length and width of the raft,~ respectively is chosen to be the  relevant control parameter. This is guided by the literature on jamming transition \cite{liu_jamming_2010,liu_nonlinear_1998}. The reported measurements are made in a range below the density where folds,~ i.e.,~ out of plane distortions \cite{vella_elasticity_2004} occur (see movie `MS2.avi' in Suppl. Information). The lack of distinct features beyond the third peak in the radial density pair-correlation function,~ $g(r)$,~ in Fig. \ref{fig.1}(d),~ illustrates that the raft remains amorphous in both the `compressed'  (see Fig. \ref{fig.1} (b)) and the `relaxed' (see Fig. \ref{fig.1} (c)) states.

\subsection{Experimental setup}

The schematic of the experimental set-up is shown in Fig. \ref{fig.1b}. The system is `compressed' (or expanded) by moving two motorized Teflon barriers along $x$ in steps of $0.05mm$.  The system is sheared sinusoidally in the $x$-direction with a hydrophobized microscope cover-slip  attached to a piezo-stage (PI-517.3CL) with a amplitude ($u_x^0 $)=0.05mm and a frequency ($\nu$)=20Hz. A stainless-steel cantilever  and a parallel plate capacitor   is used to measure  shear  ($\sigma_{xy}$)  and  longitudinal  ($\sigma_{xx}$) stresses, respectively.    Details of the experimental methods  and the parameters   used for calculating stresses  are described  in Appendix A. Additionally the system is imaged under no shear for each position of the barriers. These images are then analyzed to obtain the mean coordination number ($Z$) and to generate the Voronoi diagram  from  which    mean  cell area ($A$) is calculated.

\subsection{Measurement of mechanical response}

The mechanical response of this system to external stresses (longitudinal and shear) is described in this paper in terms of a spatially averaged effective longitudinal and shear moduli ($K_A$ and $G$,~ respectively) that are defined as: $K_A=\delta \sigma_{xx}/ \delta u_{xx}$  and $G=\sigma_{xy}/u_{xy}$,  where $u_{xy}=u_x^0/D$ is the shear strain,~ $D=60mm$  is the distance of the cantilever from the shear-launching microscope cover-slip and $\delta u_{xx}=-\delta L_x/L_{x}^0=\delta \Phi/\Phi_0$ is the incremental compressive strain,~ where  $L_x^0$  is the length of the  raft in the `relaxed' ('compressed')  state  for an compression (expansion) run  and $\Phi_0$ is the density corresponding to $L_x^0$.  Both $\sigma _{xx}$ and $\sigma _{xy}$ are measured with an instrumental resolution of 1 mPa. The noise observed in the data is intrinsic to the system and is a signature of  finite size effect of the system.  Thus  the  longitudinal modulus ($K_A$) is  calculated by  numerically differentiating   smoothed   (using a  polynomial fit) $\sigma_{xx}$ with respect to $\Phi$. We note that the elastic moduli are used here as an intuitive measure of the stress-transmission defined above \cite{hentschel_athermal_2011}.

Figure \ref{fig.2a} (a) and (b) show the variation of $\sigma_{xx}$ and $\sigma_{xy}$  with $\Phi$ for  $c1$ (open symbols) and $e1$ (filled symbols). The first compression cycle ($c1$) starts from the patchy state (see Fig. \ref{fig.1} (a))  where both $\sigma_{xx}$  and $\sigma_{xy}$  are zero,~ while the expansion cycle ($e1$) starts from the  `compressed' state (see Fig. \ref{fig.1} (c)). During the first compression,~ system-size spanning stress-bearing networks  form around   $\Phi=0.74$,~   marked by rapidly growing longitudinal and shear stresses. However,~ during the subsequent expansion cycle towards the 'relaxed' state ($\Phi \sim 0.69$,~ as in Fig. 1(c)),~ both $\sigma_{xx}$ and $\sigma_{xy}$  remain finite.

For the subsequent  cycles (the first expansion onwards) the observed variation of the stresses separates into two regions: Region I for $\Phi<0.74$,~ where $\sigma_{xy}$  decreases and region II for $\Phi>0.74$,~ where $\sigma_{xy}$  increases rapidly with $\Phi$,~ whereas $\sigma_{xx}$ increases monotonically and non-linearly with $\Phi$ in both regions.  The variations in the stress transmission characteristics  are illustrated through the computed $K_A$ and $G$,~ shown in Fig. \ref{fig.3} (a) and (b),~ respectively. The ratio,~$K_A/G$,~ shows a pronounced but inhomogeneously  jagged cusp around $\Phi=0.74$ (see Fig. \ref{fig.3}(c)) \cite{bulk}.  The transition region is broad  and  shown as a shaded region in the figures.

\subsection{Structural rearrangement }

The displacement field associated with the  particles in response to  the barrier movement is calculated by digitally cross-correlating  an image  corresponding to a specific $\Phi$ to a later one  obtained after displacing  the barriers by 0.25mm. The  details of computing the displacement field is given in  \cite{mori_introduction}.  The right panel of Fig. \ref{fig.2b}  shows the typical displacement field during the first expansion cycle. The  field is inhomogeneous and exhibits a large number of complex cellular features for $\Phi<0.74$. This is reflected in the broad distribution of the probability distribution function,~ $P(\theta)$,~  of the argument ($\theta$)  associated with the displacement vectors ($\vec {s}$̂),~ i.e.,~ $\theta =tan^{-1}⁡ (s_y/s_x )$ \cite{ellenbroek_non-affine_2009} shown in the left panel of Fig. \ref{fig.2b}. However,~ for $\Phi>0.74$  the  field becomes homogeneous and $ P(\theta)$ narrows considerably. The spatial resolution of the above  imaging method is 0.05 mm and it limits  us from calculating the displacement field associated with the shear deformation.

The inhomogeneity of the displacement field at low densities ($\Phi <0.74$) arise from  mechanical instabilities which can be interpreted as plastic  events at  the scale of the particle size (see movie `MS3.avi' and Fig. \ref{fig.2c}).  In order  to study the plasticity associated with the motion of the particles,~ each of them are individually tracked.  Representative particle displacements ($\Delta_i$)  measured from the `relaxed' state    is plotted as a function of $\sigma_{xx}$  in  Fig. \ref{fig.2c} (a) for the second compression (c2) cycle.    The jaggedness  of the graph, i.e., sudden jumps in $\Delta_i$'s,  is  associated with plasticity.  A measure of the  total number of plastic events associated with the stress cycling    is obtained from the  hysteresis of the  cumulative   strain   parameter $(\lambda),=\sum  \Delta_i /L_x^0$, where the  summation is carried out over all the particles. Figure \ref{fig.2c}(b) shows the variation  of $\lambda$ with $\sigma_{xx}$ for successive  compression (open symbols)  and expansion (filled symbols) cycles.

The images of the raft at various $\Phi$'s are used to further explore possible structural signatures of the transition. Figure \ref{fig.3c} (a) shows an increase in the mean coordination number $Z$ ($=N \int_{2a}^{2a+\epsilon} 2 \pi lg(l)dl$),~ where $\epsilon$  is chosen to match the first peak in $g(r)$ as a function of $\Phi$.  The corresponding variation  of  the Voronoi cell area $A$ normalized with the particle area ($\pi a^2$), an alternative  measure of the local structure and compactivity \cite{katgert_jamming_2010},   is  plotted in Fig. \ref{fig.3c} (b). Both $Z$ and $<A>/\pi a^2$ varies linearly with $\Phi$ in region I, but however start to deviate from the initial slope (shown by the extrapolated dotted line) in the  transition region. It suggests that the measured weakening  of the shear modulus is accompanied by structural reorganizations at the length scale of nearest neighbors throughout the system.

\section{Discussion of results}

The choice of an order parameter in a disordered system is usually not unique and is often  subjective  \cite{torquato_is_2000}. In our experiments we find the mean  coordination number $Z$ is  a  sensitive  parameter to describe the transition.  The dependence of $K_A$ and $G$ on $Z$ is shown in Fig. \ref{fig.4} (a) and (b) respectively for $e1$ and $c2$.  The system exhibits two terminal states,~ (i) `less-rigid' state ($K_A\sim 20 $Pa and $G\sim10$ Pa) and (ii) a `more-rigid' state ($K_A \sim 2000$Pa and $G \sim 200$Pa). The transition from one state to the other is marked by decrease in $G$ around $Z\sim3.6$,~ although,~ $K_A$ changes monotonically with increasing $Z$. Hence, the anomaly observed is in the variation of  the shear compliance, i.e.,  $ 1/G$,   reminiscent of the behavior of the  magnetic susceptibility at a  spin glass transition in disordered magnets \cite{binder_spin_1986}.

The cusp in  $K_A/G$ (see Fig. \ref{fig.3} (c))  is a robust feature of this transition. This is a signature of the difference in the density dependence of $K_A$ and $G$,~    reminiscent of, but distinct from,~ the power-law divergence   observed in simulations of the jamming transition in frictionless granular materials \cite{liu_jamming_2010}. However,~ one must emphasize that the phenomenology observed here is far from the original definition of athermal jamming for which only hard core interaction is considered and where the transition is between     rigid and   non-rigid states. In the conventional sense the 'relaxed' state of the  system is already in a jammed state. The  observations   should be  viewed as a transition between two jammed states whose   differences are discussed below.

\subsection{Microscopic mechanism associated with the transition}

In order to understand the microscopic mechanism associated with the transition,~ we investigated the dependence of $G$ on the particle size (see inset of Fig. \ref{fig.4} (b)). For the `less-rigid' state (filled triangles),~ $G\propto a^{-1}$. This is consistent with a capillary bridging mechanism ($G \sim \gamma/a$,~ where $\gamma$ is the surface tension of the liquid) through the pinned contact lines of the liquid on the particles \cite{kralchevsky_particles_2001,vella_elasticity_2004}. In the  `more-rigid' state  the contact friction dominates the inter-particle interaction. The measured shear modulus ($G$) scales with the number density of contacts,~ which is  proportional to the number density of particles,~ i.e.,~ $G \propto N/(L_x L_y) \sim \Phi/a^2$. It assumes that the roughness scale is independent of the particle size. Since this happens over a narrow range of $\Phi$,~ one obtains $G\propto a^{-2}$ as in conventional elasticity in a quasi-2D system \cite{chaikin_principles_2000}. We hence conclude that regions I and II represent a capillary-bridged solid and a frictional solid,~ respectively.

\subsection{Time dependent effects and stress cycling in the particle raft}

The structural relaxation time of the particle raft,~ $\tau=12\pi \eta a^3/mv^2 \sim 1000s$,~ where $ m=0.2 \mu g$ and $v=50 \mu m/s$ are the mass and velocity of a single particle respectively and $\eta=1mPas$ is the viscosity of the liquid. In calculating the relaxation time,~ we assume that the effective temperature of the system ($T_{eff}$) is related to the kinetic energy of the particle,~ i.e.,~ $T_{eff}=mv^2/2K_B$,~ where $K_B$ is the Boltzmann's constant. The system thus takes a long time to reach its equilibrium state. This would mean that the system would typically find itself in a kinetically arrested state and would  show a    dependence on the history of  the paths via which a given state of the system is reached. Indeed, measurements above 1 mHz probes a frequency-dependent rigidity of the material as shown in  Fig. \ref{fig.5}(a) for $\Phi=0.77$.  Moreover,~ creep effects (time-dependence) in $\sigma_{xy}$  are more pronounced in region II than in region I,~ as shown in   Fig. \ref{fig.5} (b). Interestingly,~ the stress allows the `more-rigid' frictional solid to explore various metastable states and hence the system shows a marked `creep',~ i.e. a temporal variation,~ in $\sigma_{xy}$.  This is absent for the `less-rigid' capillary solid suggesting that it is closer to a deeper metastable minimum. Strong  history dependence is also seen in  the variation of $\sigma_{xx}$ and $\sigma_{xy}$ measured for subsequent compression and expansion cycles  which broadly follow the trend (see fig. \ref{fig.2a})  but with greatly reduced hysteresis (see Fig. \ref{fig.hyst}, the data shown in the figure is for  particles whose diameter is 1mm),~ analogous to residual densification observed in amorphous materials \cite{loerting_multiple_2009}.

\section{Conclusion}
We have shown that a  rigid particle raft  can undergo a  phase transition from a  `less-rigid'  low density state to a `more-rigid' high density state as a function of particulate number density. The transition is marked by a weakening of the shear modulus.  The measured shear modulus which distinguishes the two states of the system,~ i.e.,~ `less-rigid' and `more-rigid',~ arises from the restoring force of pinned contact lines in the former case and from particle contacts through a frictional coupling in the latter.  The weakening of the shear modulus observed in the crossover region is thus attributable to a reduction of restoring force arising from plastic events caused by the mechanical instabilities and the associated depinning of the contact lines \cite{bourne_elastic_1986,coppersmith_phase_1990}.

Although the results presented in the present paper are specific to  particle rafts,~ the weakening of the shear response   and local reorganization  seen in the present experiment,~   have  also been  observed in network glasses \cite{thorpe_rigidity_1999,poole_polymorphic_1997}.  We hope that the present experiment will  provide some insight into the complex phenomenon of  pressure induced phase transitions in amorphous solids
\cite{pham_multiple_2002,mishima_apparently_1985,poole_amorphous_1995,sun_pressure-induced_2011,loerting_multiple_2009,mishima_relationship_1998}.

The authors thank M. Bandi, M. Cates, N. Menon, L. Mahadevan, S. R. Nagel and A. Yethiraj for helpful discussions.

\appendix
\section{Experimental set-up}

\begin{figure}
\includegraphics[width=0.75\textwidth]{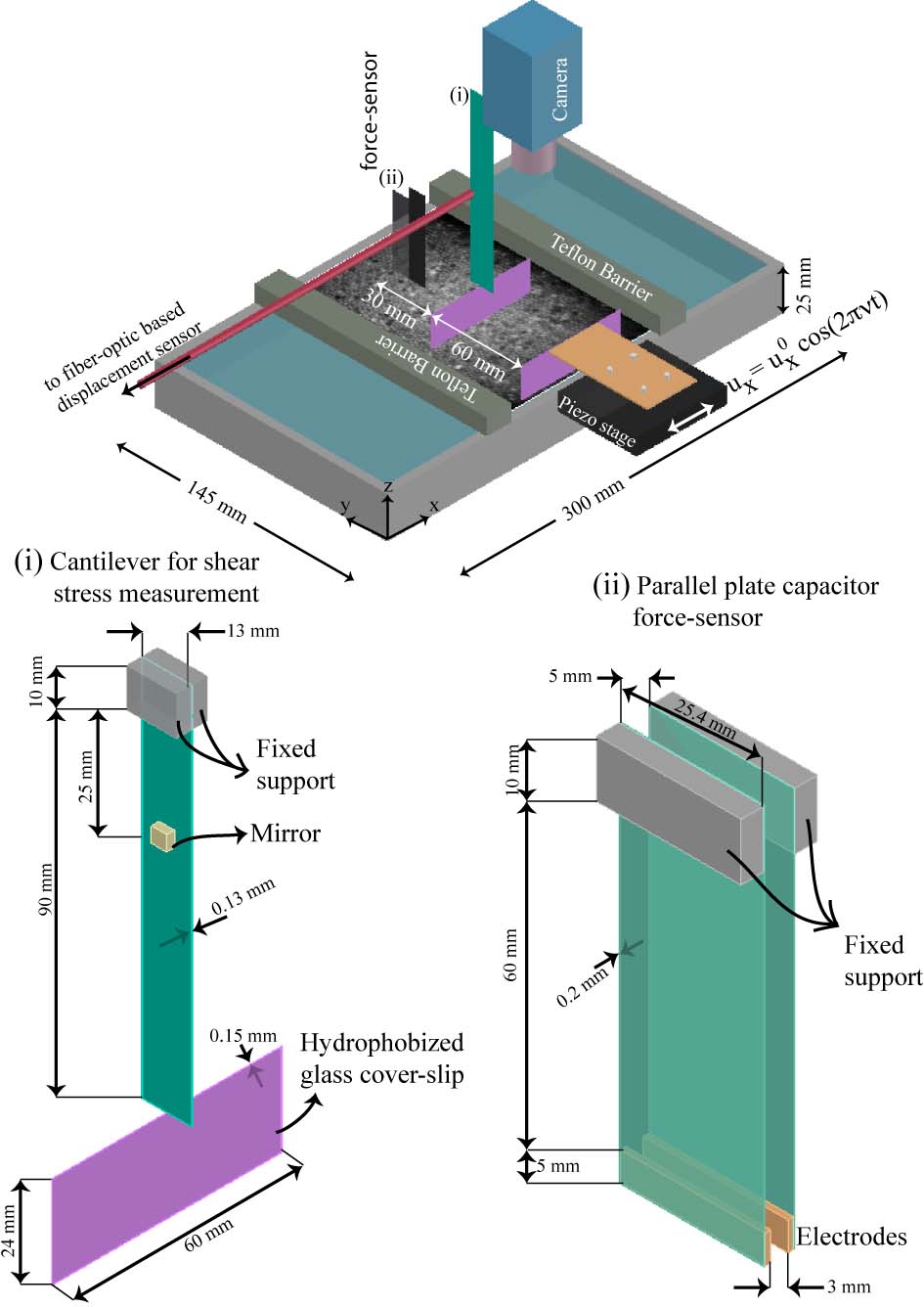}
\caption{Experimental setup: hydrophobized microscope cover-slip attached to a piezo-stage oscillates along $x$ producing a sinusoidal shear deformation. The cantilever with an attached microscope cover-slip,~ placed along $y$ at a distance ($D$)=60mm away from the piezo-stage measures the shear stress ($\sigma_{xy}$). A parallel-plate capacitor,~ marked as the force-sensor,~ is used to measure the longitudinal stress ($\sigma_{xx})$. }
\label{fig.1b}
\end{figure}

\begin{figure}
\includegraphics[width=0.75\textwidth]{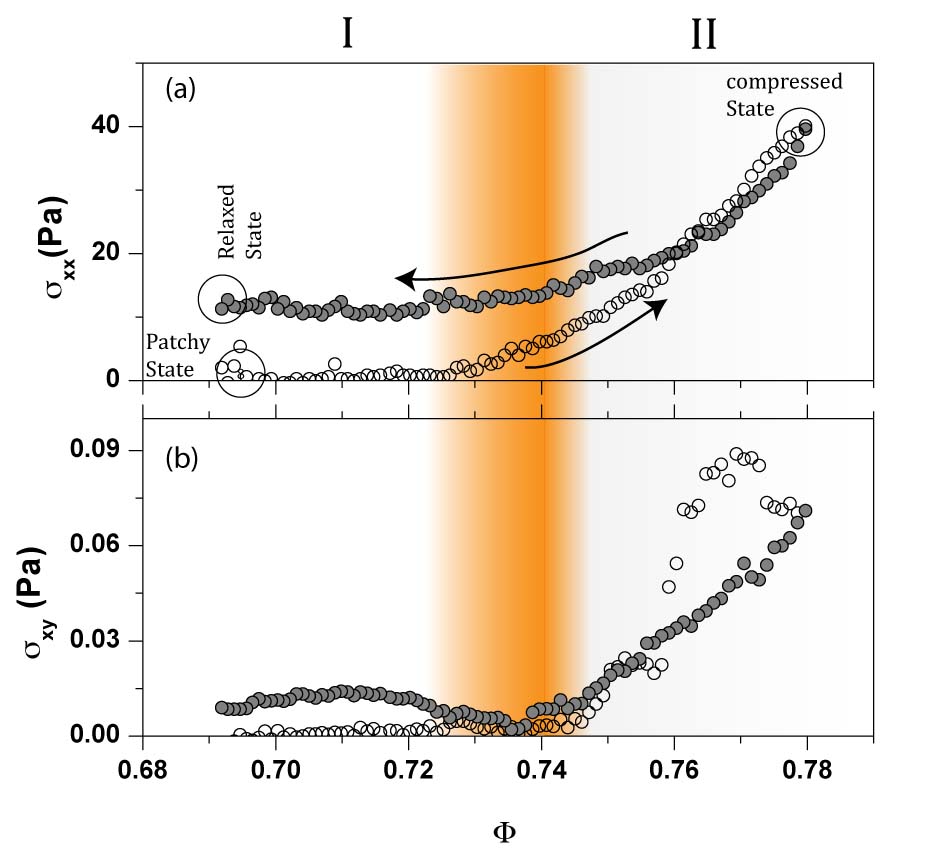}
\caption{ Variation of longitudinal stress ($\sigma_{xx}$) and shear stress ($\sigma_{xy}$) are plotted as a function of $\Phi$ in (a) and (b) respectively for  $c1$ (open symbols) and $e1$ (filled symbols) cycles. The arrows in (a) shows the direction in which $\Phi$ changes. The region over which $\sigma_{xy}$ shows a softening is marked as region I. In region II,~ $\sigma_{xy}$ increases rapidly. The different states of the system i.e.,~ patchy,~ `compressed' and `relaxed' are marked in the figure. The  transition region separating I and II is shaded.
 }
\label{fig.2a}
\end{figure}

\begin{figure}
\includegraphics[width=0.75\textwidth]{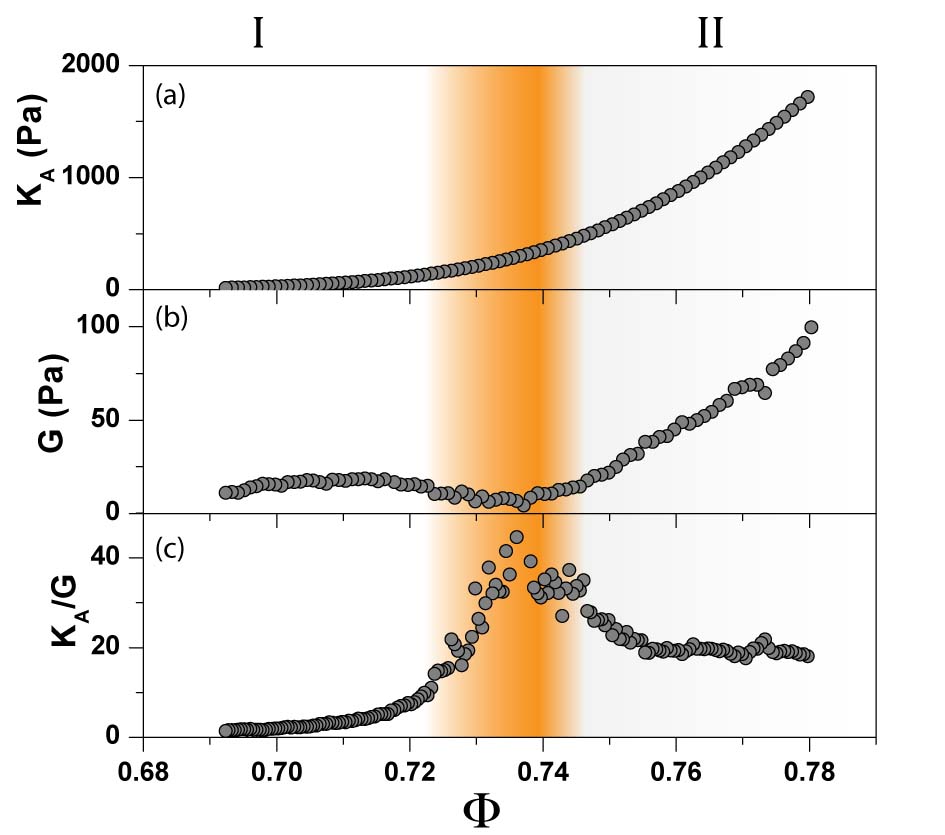}
\caption{The variation of (a) longitudinal modulus ($K_A$),~ (b) shear modulus ($G$),~ (c) and  the ratio $K_A/G$  as a function of $\Phi$ for the first expansion cycle.The longitudinal modulus ($K_A$) is calculated by  numerically differentiating   smoothed   (using a  polynomial fit) $\sigma_{xx}$ with respect to $\Phi$.  }
\label{fig.3}
\end{figure}

\begin{figure}
\includegraphics[width=0.75\textwidth]{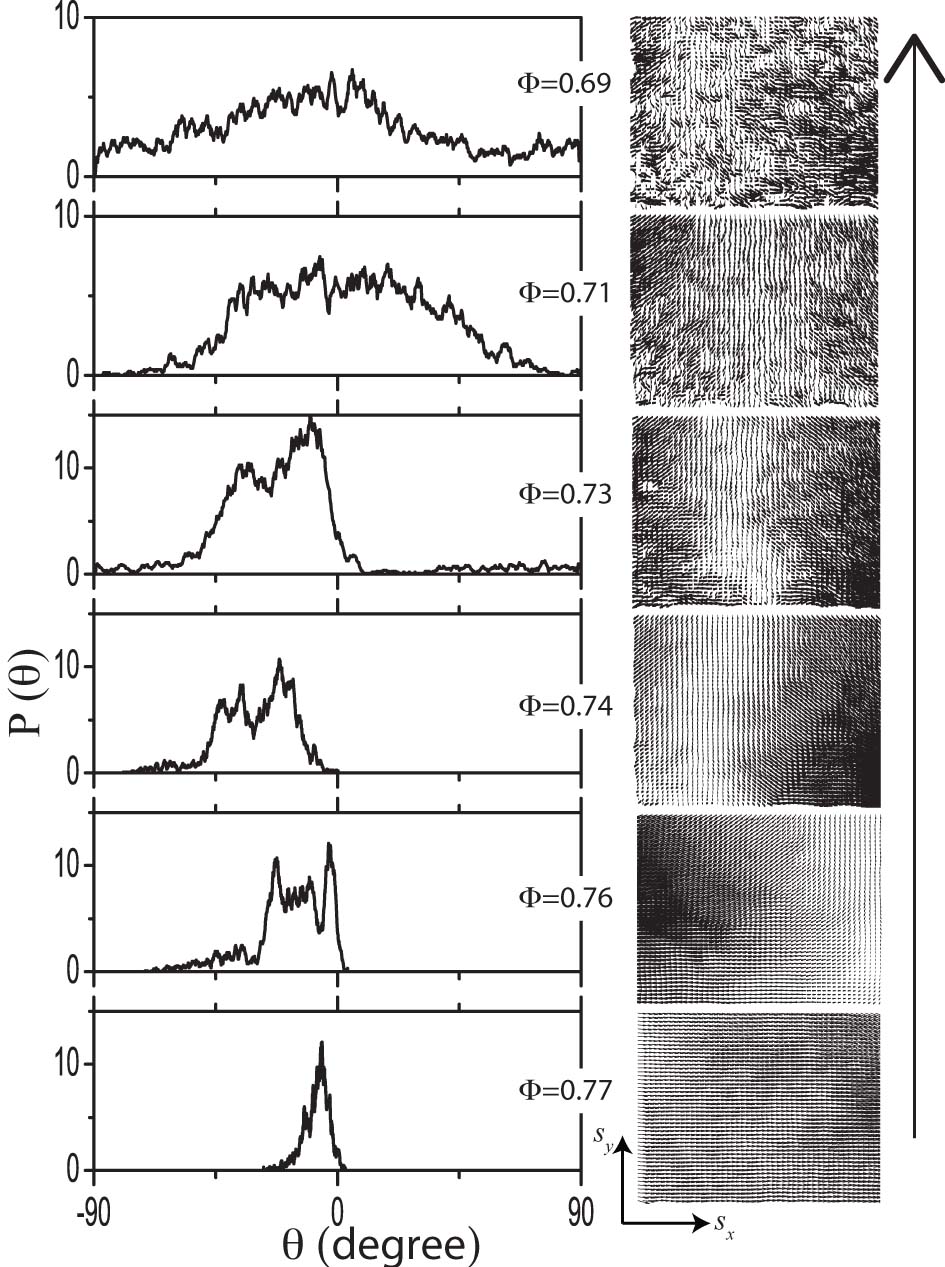}
\caption{Displacement field  of the particles  and the  probability distribution of angles,~ $P(\theta)$ associated with the displacement vector ($\vec{s}$) measured during the first expansion cycle is shown in right  and left panel respectively.  Here $\theta=tan^{-1}(\frac{s_y}{s_x})$ For values of $\Phi<0.74$ the displacement field is inhomogeneous and exhibits a large number of complex cellular features (right panel). This is reflected in the   broad distribution of   $P(\theta)$   (left panel). However,~ for values of $\Phi>0.74$,~ the displacement field becomes spatially correlated over large distances and  $P(\theta)$ narrows considerably. The arrow indicates the progressive sequence of $\Phi$'s, i.e, the system  moves from a state of high density, i.e., $\Phi=0.77$ to a state of low density, i.e., $\Phi=0.69$.
 }
\label{fig.2b}
\end{figure}

\begin{figure}
\includegraphics [width=0.75\textwidth]{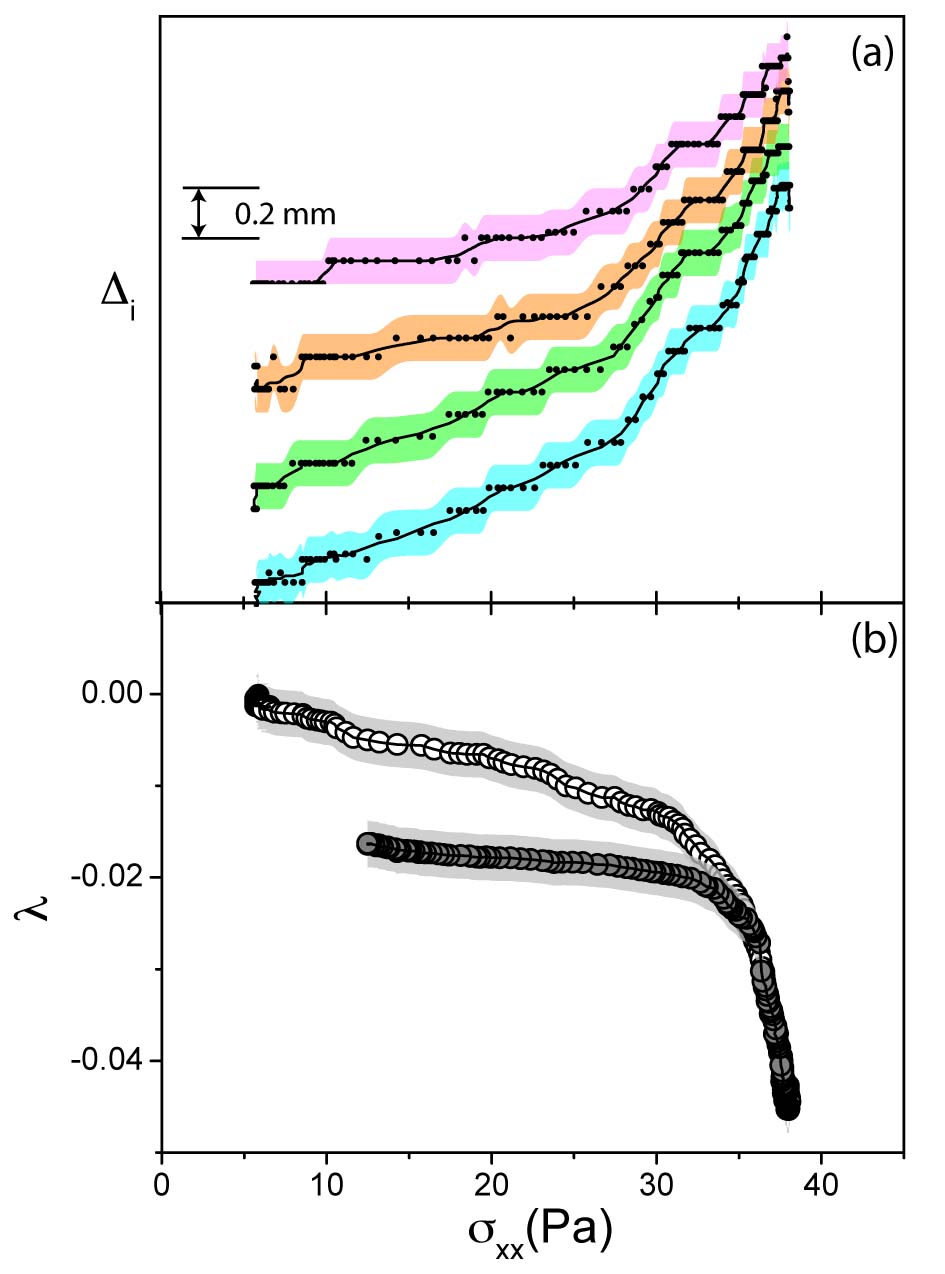}
\caption{  (a)Typical displacement curves of a few   particles ($\Delta_i$),~  measured from its position in the `relaxed' state as a function of $\sigma_{xx}$.  For visual clarity the curves are  shifted vertically.  The corresponding scale bar of  the displacement is shown in the figure. The data is plotted for a compression cycle.     (b) The variation of the cumulative  strain ( $\lambda=\sum  \Delta_i /L_x^0$, where the summation is over all the particles) with $\sigma_{xx}$ for successive compression (open symbols) and  expansion (filled symbols) cycles. The uncertainty in the measurement of $\Delta_i$ and $\lambda$ is shown by the  accompanying band.
}
\label{fig.2c}
\end{figure}

\begin{figure}
\includegraphics[width=0.75\textwidth]{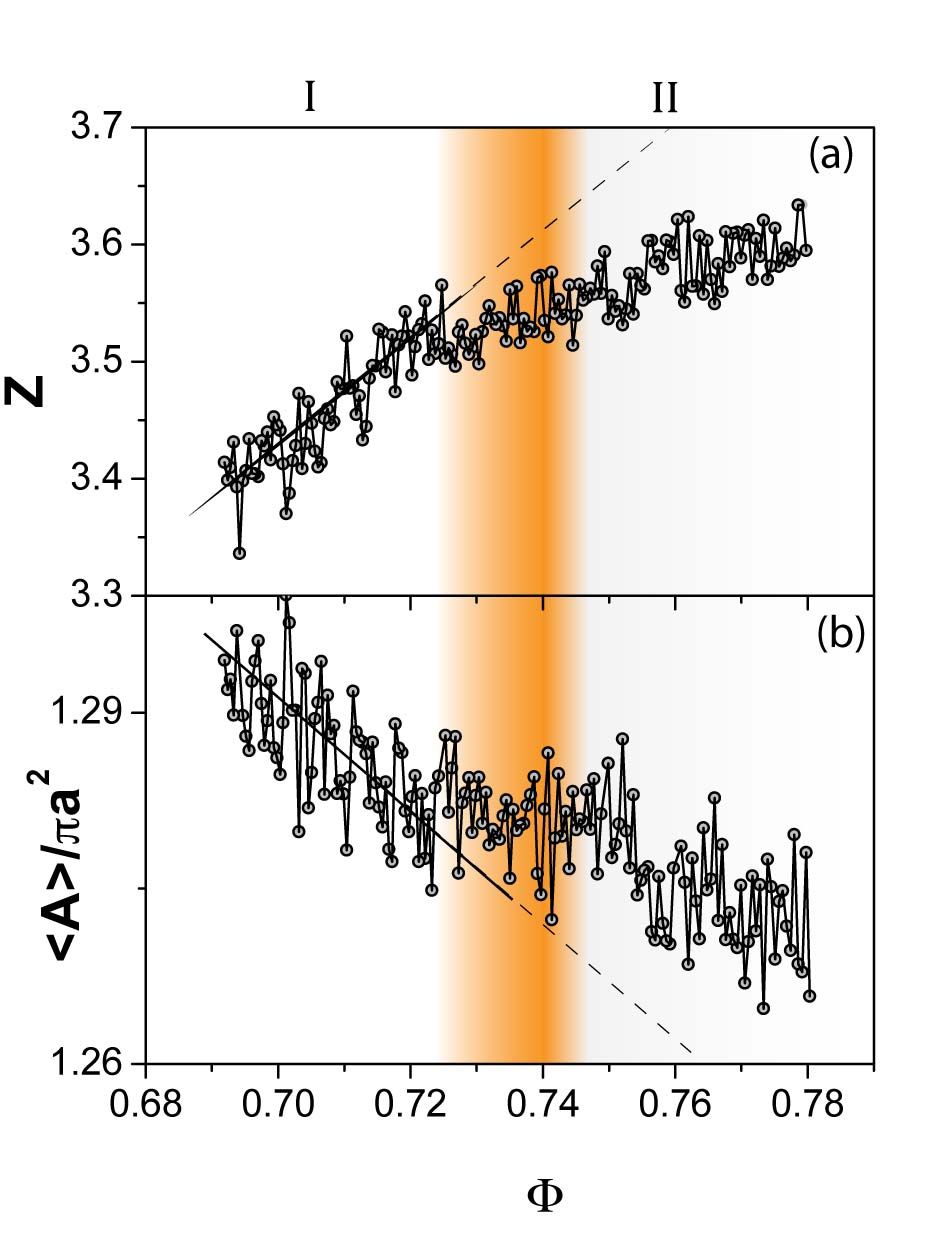}
\caption{The variation of  (a) the  mean coordination number $Z$    and   (b) the mean  Voronoi cell area $A$ normalized with the particle's  area ($\pi a^2$) as a function of $\Phi$. The solid lines  show their linear variation  in region I.  The dotted lines  which  are a linear extrapolation of initial variation  highlights the later deviation of the graphs.   The data presented here is for  $e1$.
}
\label{fig.3c}
\end{figure}

\begin{figure}
\includegraphics[width=0.75\textwidth]{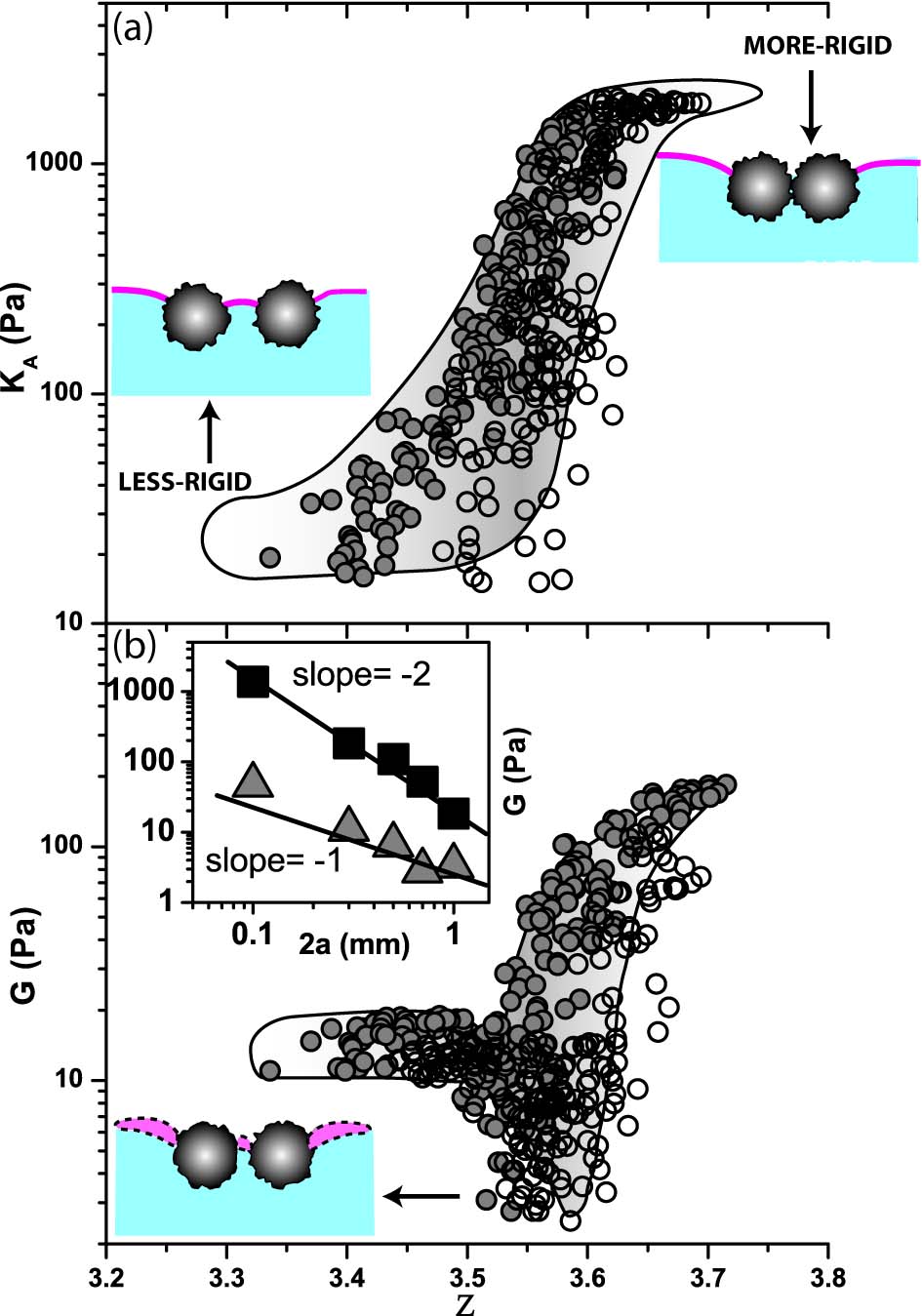}
\caption{Variation of (a) $K_A$ and (b) $G$ as a function of $Z$. The filled  symbols correspond to  $e1$ while the open ones correspond to  $c2$. The shaded regions indicate the spread in the data. The `less-rigid' (schematically shown as a capillary bridged interaction in (a)) and `more-rigid' states (schematically shown as a friction dominated interaction in (a)) are marked in the figure. The inverted cusp in $G$ versus $ Z$  corresponds to the slipping of the contact line (schematically shown in (b) with a greater width of a sliding contact line). The inset of (b) shows the variation of $G$ of the `less-rigid' (triangles) and `more-rigid' (square) states of the raft as a function of particle diameter ($2a$). The solid lines passing through the data show $a^{-1}$ and $a^{-2}$ variation of $G$ for the `less-rigid' and `more-rigid' states,~ respectively. }
\label{fig.4}
\end{figure}

\begin{figure}
\includegraphics[width=0.75\textwidth]{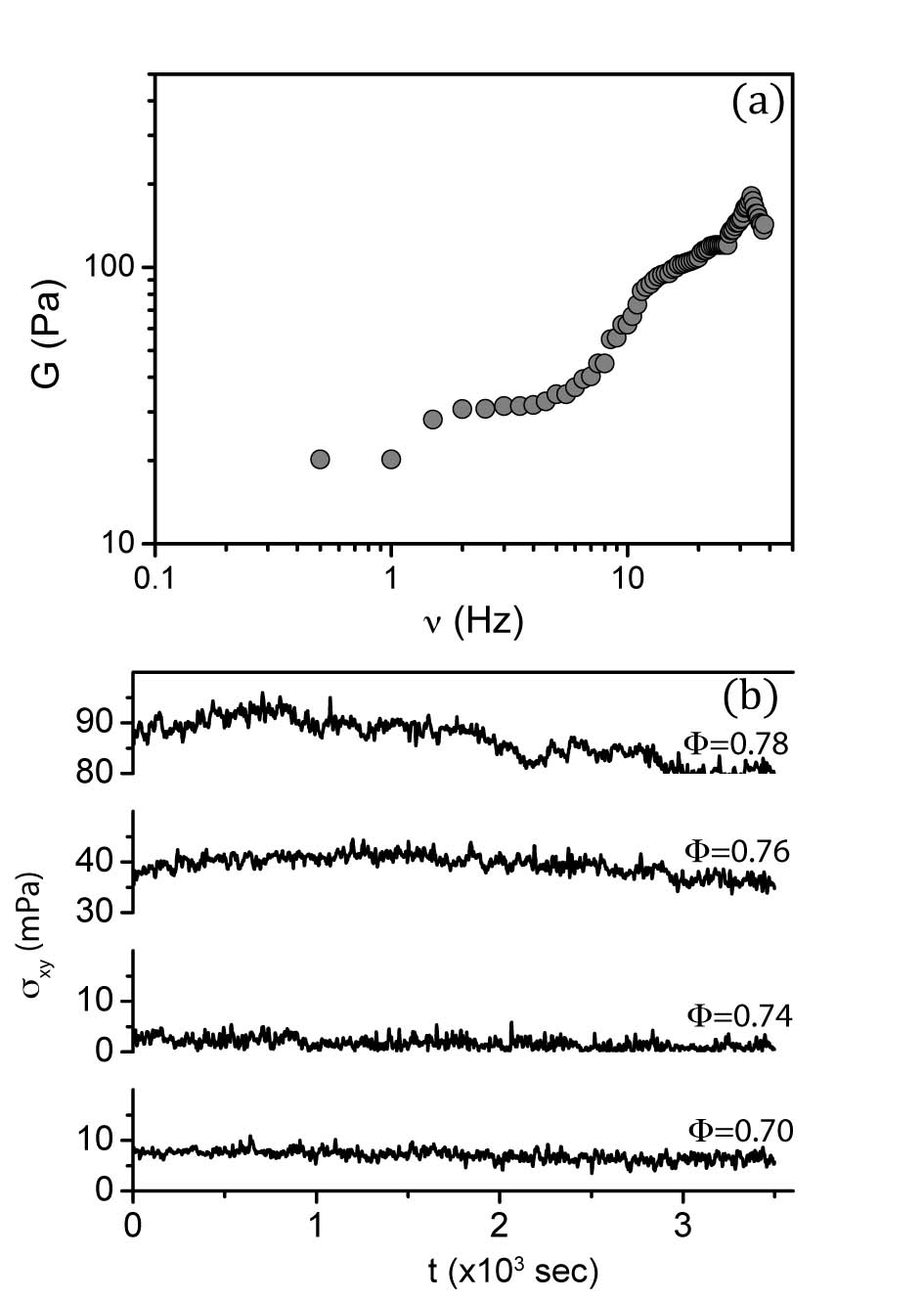}
\caption{(a) Frequency dependence of $G$ measured at $\Phi=0.77$.  (b) The time dependence  (creep effect)of $\sigma_{xy}$ is shown for  different $\Phi$'s. }
%0.70,~ 0.74,~ 0.76 and 0.78}
\label{fig.5}
\end{figure}

\begin{figure}
\includegraphics[width=0.75\textwidth]{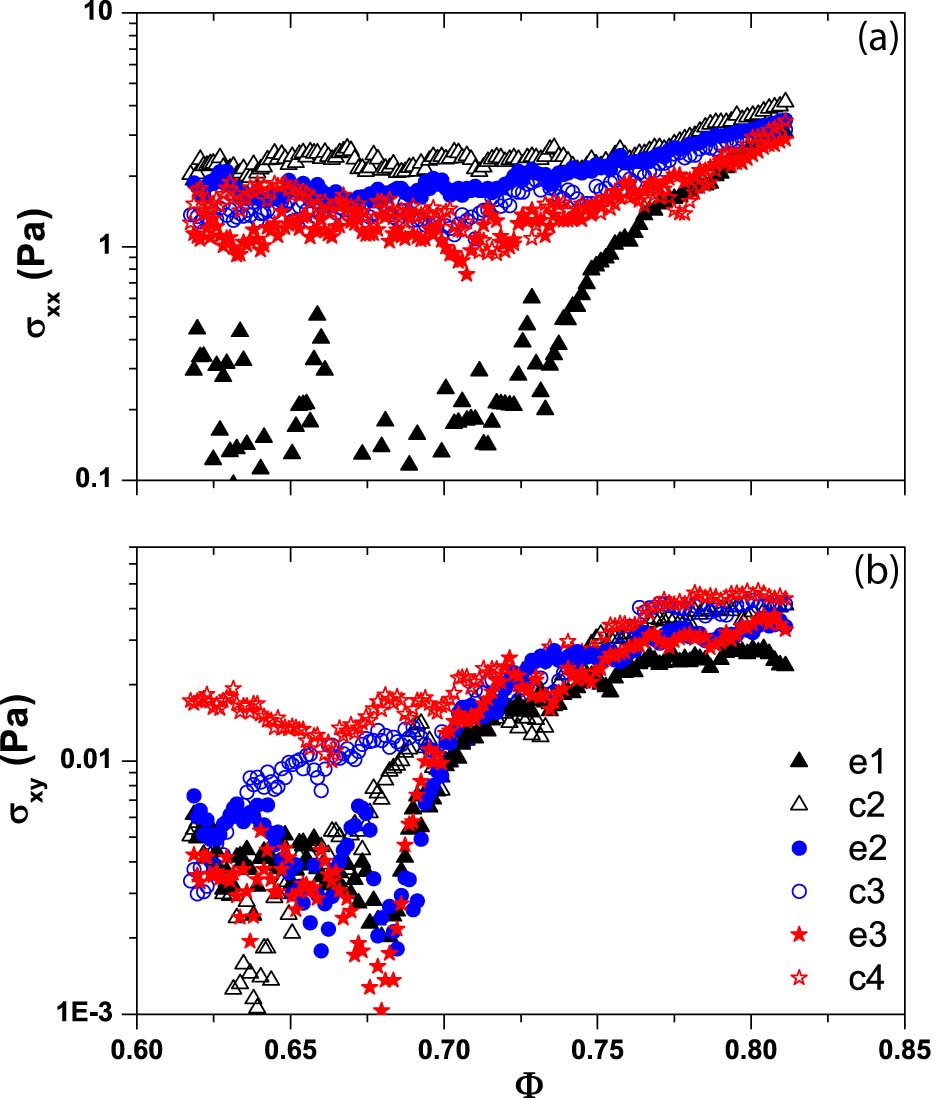}
\caption{ The figure shows the variation of $\sigma_{xx}$ and$\sigma_{xy}$ for various compression (open symbols) and expansion (filled symbols) cycles as a function of $\Phi$ for particles of diameter 1 mm.  The sequence of expansion and compression are as follows,~ first compression (data not shown for reasons discussed in the main text) : first expansion ($e1$: filled triangles),~ second compression ($c2$: open triangles),~  second expansion ($e2$: filled circles),  third compression ($c3$: open circles),  third expansion ($e3$: filled stars),~  fourth compression ($c4$: open stars). The repeated compression-expansion cycles result in annealing of the raft and thus later cycles show a reduced but finite hysteresis. }
\label{fig.hyst}
\end{figure}

The schematic of the experimental set-up is shown in Fig. \ref{fig.1b}. Hydrophobic silica particles are sprinkled on the air-water interface in a Langmuir trough (300mm x 145mm x 25mm) made of Teflon. These particles coalesce and form disjoint (patchy) clusters. The two motorized Teflon barriers of the trough are used to fuse these clusters into a homogeneous `compressed' or 'relaxed' state by moving the barriers inward or outward in $x$-direction respectively. The system is illuminated uniformly from the bottom using an LCD monitor through a glass window sealed in the trough and a camera above records the image in the $(x-y)$ plane at each barrier step ($=0.05mm$).  The system is sheared in the $x$-direction by moving a hydrophobic cover-slip (60mm x 24mm x 0.15mm) sinusoidally $u_x=u^0_x{\rm cos}(2\pi \upsilon t)$,~ where $u^0_x=0.05mm$ and frequency $\upsilon =20Hz$,~ connected to a three-axis piezo-stage (PI-517.3CL).  A stainless steel cantilever is kept at $D=60mm$ in the $y$-direction from the shear launching cover-slip as marked by (i) and shown at the left-bottom of the Fig. \ref{fig.1b} with dimensions mentioned. The hydrophobic cover-slips attached to piezo stage and the cantilever makes contact with the particles at the air-water interface. The shear stress (${\sigma}_{xy}$) is measured from the lateral displacement of the cantilever. The displacement is sensed by a fiber optic-based displacement sensor (MTI-2000 FOTONIC) from the interference of light that is reflected from the mirror glued on the cantilever at a distance of $25mm$ from the fixed support.  The output of the displacement sensor  is fed to a lock-in amplifier (SR830).  A parallel plate capacitor,~ used as a force-sensor,~ is marked by (ii) and also shown with dimensions in the right-bottom of Fig. \ref{fig.1b}. It is made of $0.2\ mm$ thick flexible polymeric sheet and kept at a distance of  $30mm$  from the cantilever in the $y$-direction. The metal electrodes are glued at the bottom of the two plates of the force-sensor with a separation of 3mm and are immersed in water. A capacitance-bridge (1615-A General Radio) is used to detect the change in the plate separation during compression (or expansion) from the lateral deflection of the individual capacitor plates. The output of the capacitance-bridge is fed to another lock-in amplifier (SR830) which drives the bridge at 100KHz. The stresses,~ shear (${\sigma}_{xy}$) and longitudinal ($\sigma_{xx}$),~ are calculated using ``bending-of-a-beam'' formula using lateral displacement of the cantilever in the former case and the plate deflection in the latter.  The formula used in general is given by: the stress $\sigma =6REI/x^2_0(3l_0-x_0)\alpha $ \cite{landau_theory_1986} where $R$ is the lateral displacement,~ $E$ is Young's modulus,~ $I=bh^3/12$ is the second moment of area ($b$ and $h$ are width and thickness of the detecting object,~ respectively),~ $l_0$ is the length of the object,~ $x_0$ is point of detection from the fixed end and $\alpha $ is an effective area of contact with the particles.  All these parameters used in the calculation of stress' are tabulated in table \ref{table:1} for the cantilever and the parallel plate force sensor.

\begin{widetext}
\begin{table}
\begin{tabular}{|p{0.9in}|p{0.9in}|p{0.9in}|p{0.9in}|p{0.9in}|p{0.9in}|p{0.9in}|} \hline
 & length ($l_0$) (mm) & width ($b$) (mm) & thickness ($h$) (mm) & Young's modulus ($E$) (GPa) & point of detection ($x_0$) (mm) & effective area ($\alpha$)\newline (mm${}^{2}$) \\ \hline
Cantilever & 90 & 13 & 0.13 & 200 & 25 & 30 \\ \hline
Parallel plate force-sensor &  60 &  25.4 &  0.2 &  2 &  60 &  12.7 \\ \hline
\end{tabular}
\caption{Parameters used in the calculation  of $\sigma_{xx}$ and $\sigma_{xy}$ for the cantilever and the parallel plate force sensor}
\label{table:1}
\end{table}

\end{widetext}

%%\bibliography{myrefs}	
%\input{pre_Final_1_bbl.tex}

%\bibliographystyle{natbib}
%\bibliography{myrefs}	

\begin{thebibliography}{33}
\bibitem[{cap()}]{capillary}
$l_c=\sqrt{(\gamma/\rho g)}\sim 2mm$, where
  $\gamma=70mN/m$, $\rho=1000kg/m^3$ and $g=9.8m/s^2$ are the surface tension
  of water, density of water and acceleration due to gravity respectively.
\bibitem[{bul()}]{bulk}
In continuum elasticity,~ the bulk modulus $B=K_A-4G/3$ ;
  here,~ $ K_A \gg G$,~ thus $ B\sim K_A$.
\end{thebibliography}
%\input{biblio.tex}
%\end{thebibliography}

\end{document}